# A social license for nuclear technologies

Seth Hoedl, Ph.D., J.D.[*]

**Abstract**  Nuclear energy technologies have the potential to help mitigate climate change.  However, these technologies face many challenges, including high costs, societal concern and opposition, and health, safety, environmental and proliferation risks.  Many companies and academic research groups are pursuing advanced designs, both fission and fusion-based, to address both costs and these risks.  This Chapter complements these efforts by analyzing how nuclear technologies can address societal concerns through the acquisition of a social license, a nebulous concept that represents "society's consent" and that has been used to facilitate and improve a wide range of publicly and privately funded projects and activities subject to a range of regulatory oversight, including large industrial facilities, controversial genetic engineering research, and environmental management. Suggestions for public engagement and consent-based siting, two aspects of a social license, have been made before. This chapter modernizes these suggestions by briefly reviewing the social license and engagement literature.  The Chapter discusses, in the context of how to acquire a social license, the role of government regulation, the role of project proponents and government actors, and the role of four key principles, including engendering trust, transparency, meaningful public engagement, and protection of health, safety and the environment. Further, the Chapter uses the social license concept to explain why some nuclear waste repositories have succeeded while others languish and provides concrete recommendations for the deployment of new nuclear waste repositories and advanced power plants, both fission and fusion-based.

**Contents**

---

[*]Chief Operating & Science Officer, Post Road Foundation.

S Hoedl (✉), 1935 Addison St. STE A, Berkeley, CA 94704 USA, shoedl@gmail.com

# #.1 Introduction

The benefits of nuclear technologies are well known and discussed elsewhere in this volume. However, nuclear energy technologies can be difficult to implement. They are often expensive, generate radioactive waste, and are inevitably connected to nuclear weapons, either directly through the materials and facilities used, or through knowledge transfer that can be redirected for weapons purposes. As a consequence, nuclear energy technologies have long been controversial. Power plants, waste repositories, enrichment and fuel processing facilities have long been met by legal and/or civil protest.[1] Opposition is driven by, *inter alia*, concerns regarding capital expense, radiation exposure, radiation leaks, catastrophic accidents, waste stewardship, non-proliferation, terrorism,[2] and fears that nuclear power exposes individuals to risks that they cannot control.[3]

Despite the opposition to nuclear technologies, there is a need for new facilities associated with electric power production. Regardless of nuclear power's future, waste repositories need to be built and operated to deal with nuclear wastes that have accumulated since the dawn of the nuclear age. More controversially, there may also be a need for new nuclear power plants. Multiple studies suggest that humanity is unlikely to meet the climate change challenge without at least some role for nuclear power.[4] More speculative nuclear technologies, such as fusion-based power plants, may play an even larger role than fission-based power plants due to inherent safety and waste advantages.

The past challenges of building new nuclear waste repositories and fission-based power plants are well described in the literature.[5] How to proceed in the future remains debated. Likely, many different strategies are needed. Academics, companies and start-ups are actively pursuing new fission-based and fusion-based designs, with the intent of lowering costs and health, safety, environmental and proliferation risks.[6] Others are pursuing regulatory reform to stream-line the legal licensing and approval process in order to lower development and deployment costs.[7] Yet, a risk-reducing technical solution is unlikely sufficient to address societal concerns, as they are not based on quantifiable risks.[8] For example, despite the fact that coal mining and fossil fuel combustion arguably harm more people than nuclear,[9] they do not elicit the same level of opposition. Strategies other than technology development will be needed to address these concerns.

---

[1] For a review of early opposition in the 1970s, when nuclear power was expanding rapidly, see Bickerstaffe and Pearce 1980; Falk 1982. Such challenges and protests continue to occur worldwide. See The Economist 2013.

[2] Bickerstaffe and Pearce 1980, pp 313–320.

[3] Mufson 1982, p 60., citing Otway et al. 1978.

[4] Both the International Energy Agency and the IPCC report that nuclear power production should at least double over the next thirty years in order for the earth's temperature rise to remain below the 2˚C goal of the Paris Agreement. Edenhofer 2014; International Energy Agency 2015.

[5] Bickerstaffe and Pearce 1980; Slovic et al. 1991; Macfarlane and Ewing 2006.

One long recommended strategy to address concerns for nuclear projects is to adopt processes by which projects are approved or "sited" through some kind of public engagement.[10] However, the history of engagement is mixed and its ability to address concerns is greatly dependent on the nature of the engagement process. While engagement has been successful in many contexts, as described below, there have also been failures that have led to disillusionment, by both the public and project proponents. Early efforts in using engagement in the nuclear context were, in some cases, counter-productive and led to the cancellation of projects in which billions of dollars had already been invested.[11] Commentators suggest that poorly run engagement can become a "talking shop" that creates ambiguities, delays decisive action, and breads cynicism.[12] Engagement can also be coopted so as to push aside already marginalized groups.[13] Nevertheless, many lessons have been learned regarding effective engagement in fields other than nuclear, such as biomedical research, industrial facilities, and environmental management. Best engagement practices have emerged.[14]

---

[6] Lassiter 2018. http://news.mit.edu/2018/mit-newly-formed-company-launch-novel-approach-fusion-power-0309; Sorbom et al. 2015.

[7] Nuclear Energy Institute 2018.

[8] Otway et al. 1978. It has been long known that individuals ascribe more risk to radiation than experts ascribe, and more risk than other hazards. This asymmetry, also generated by other hazard types, has been explained by the fact that radiation risk is perceived to be, *inter alia*, less voluntary, more catastrophic, and more likely to be fatal than other hazards. Slovic 1987; Slovic 1996. Further, the fact that radioactive material is invisible instills a sense of dread regarding nuclear accidents – radioactive accidents have no visible end. Erikson 1990.

[9] Coal combustion alone is estimated to kill 366,000 people in China per year. Wong 2016. In contrast, the Chernobyl accident is expected to result in only 4,000 deaths in total. International Atomic Energy Agency 2008.

[10] Bickerstaffe and Pearce 1980, p 326.

[11] For example, Austria held a two-year long public debate and referendum regarding placing a completed nuclear plant into operation between 1976 and 1978. At the conclusion of the debate, voters rejected the plant. Mufson 1982, pp 55–57.

[12] Reed 2008, p 2421.

[13] Reed 2008, p 2420.

This Chapter aims to briefly translate these best practices to the nuclear context and thereby give recommendations to technologists and policy makers pursuing nuclear waste siting and advanced fission or fusion-based reactors. Although many of these recommendations have been made before,[15] a renewed look is timely for at least three reasons. *First,* recommendations for nuclear projects should be based on recent engagement research. *Second*, as mentioned above, technologists and policy makers have a renewed interest in nuclear technologies for climate change mitigation. There is an opportunity to inform their efforts. *Lastly*, many longstanding engagement recommendations have yet to be implemented, especially with regards to nuclear waste disposal in the U.S.[16] A discussion grounded on recent engagement research contributes to the ongoing debate.

This Chapter approaches engagement through the concept of a social license, a nebulous concept that represents "society's consent" to a particular project or endeavor.[17] Under this approach, engagement, legal licensing, and other project development processes are collectively conceived of as contributing to the issuance of a social license. Note that a social license is very different than a legal license. While a legal license is granted through an established and formal procedure and typically memorialized through written permits, a social license is not formally granted and is not written. Further, while a social license may require a legal license, the converse is generally not true. In fact, because of the nebulous nature of a social license, project proponents are often not aware of whether they have been granted a social license until they either lose it or are confronted with the fact that they do not have it, through formal legal action, civil protest or other means.[18]

---

[14] Reed 2008, pp 2418, 2421; Reed et al. 2018.

[15] Slovic et al. 1991, pp 1606–1607.

[16] For example, some commentators recommend nearly twenty years ago that nuclear waste repositories be sited through a consent-based process. Slovic et al. 1991, p 1607., *citing* E. R. Frech, in Proceedings of the 1991 International High-Level Radioactive Waste Management Conference (American Nuclear Society, La Grange, IL, 1991), vol. 1, pp. 442-446. Yet, as further discussed below, present U.S. law does not use a consent-based process.

[17] Note that the terms a "social license" and a "license to operate" are often used interchangeable in the literature. Both terms are often used to describe all obligations that an activity must meet, including legal, economic and social. Gunningham et al. 2004, p 329. This chapter adopts this meaning for a "social license," so that a social license can equal or exceed legal obligations. Other authors interpret a "social license" as necessarily separate and distinct from legal obligations. Bankes 2015; Canadian Association of Petroleum Landmen 2017. For these authors, a "social license" refers only to obligations that go beyond the requirements of statutes, regulations, permits or treaties. *Id*. What constitutes "society" and "consent" are context-dependent questions that are further discussed below.

The remainder of the Chapter is organized into three sections: (i) an overview of the social license concept; (ii) an illustration of the social license concept through two nuclear waste repository case studies; and (iii) specific recommendations that would help facilitate a social license for nuclear waste repositories and advanced fission and fusion-based power plants.

## #.2 The Social License

### #.2.1 Background

The social license concept has been applied to a diverse set of endeavors funded by public and private sources, and subject to a wide range of regulatory oversight, including both voluntary codes of conduct and the full panoply of environmental regulations. The social license concept explains why some projects are ultimately built while others languish and provides recommendations to proponents seeking to build new facilities. It has been applied to extractive and other energy projects, such as mining, carbon capture and sequestration, and wind farms,[19] medical[20] and other scientific research,[21] genetic engineering,[22] gain-of-function biological research,[23] experimental ecological intervention,[24] and environmental management generally.[25] In the context of industrial facilities, it has been defined as "the demands on and expectations for an [activity] that emerge from neighborhoods, environmental groups, community members, and other elements of the surrounding civil society."[26] Crucially, a social license pertains to more than just strict environmental and economic harms and benefits – it also considers societal and political values.

---

[18] Rooney et al. 2014, p 209.

[19] Hall et al. 2015.

[20] Dixon-Woods and Ashcroft 2008; Carter et al. 2015.

[21] Raman and Mohr 2014.

[22] National Academies of Sciences, Engineering, and Medicine 2016.

[23] National Science Advisory Board for Biosecurity 2016. Although the exact details of some types of gain-of-function research is not public due to dual use concerns, the governing regulations were developed through public engagement and explicitly considered ethical values other than technical risks.

[24] For example, an experimental ecology project in Ontario, Canada, the Experimental Lakes Area, pays particular attention to societal concerns. See https://www.iisd.org/.

[25] Note that the environmental management and conservation literature generally does not use the language of social license but has developed many similar concepts in the context of public engagement. Fiorino 1990; Reed 2008; Reed et al. 2018.

[26] Gunningham et al. 2004, p 308.

The social license concept emphasizes process over outcomes.[27] A social license is most likely to be granted and maintained, and projects are more likely to be completed, when community concerns are addressed in a meaningful process that opens expertise "to new questions and perspectives" rather than "letting people see the experts at work."[28] This process is more "bottom-up" than "top-down"; it is distinctly different than an expert-led technocratic approach to public concerns.[29] Critically, proponents must consider what people actually worry about instead of what they should worry about. Societal concerns must be meaningfully addressed: they cannot be dismissed out of hand by proponents or experts. Rather, the processes of developing, approving or undertaking a project must be such that societal concerns are thoughtfully considered and analyzed.

This process does not mean that all concerns of all stakeholders are always placated. However, the process itself is a powerful method to acquire consent, even from those who may disagree with the ultimate outcome, by creating a sense of "procedural justice" whereby "people affected by decisions" are able to "participate in making" such decisions.[30] A process of "good-faith efforts toward respectful listening, creative compromise, and flexible practice"[31] increases the legitimacy of a final decision[32] and lessens opposition. A meaningful process also fulfills a normative need to give citizens an opportunity to participate in decisions that affect themselves and their communities.[33]

A concrete example of how a bottom-up rather than a top-down process works in practice is provided by pulp mill factories that changed the order of their planning processes in order to facilitate a social license. Before the factories adopted a social license approach, when considering an expansion or renovation, the factories would develop a plan, hire engineers to design the new facility, apply for necessary environmental permits, and then present the plan and permits to the community as a *fait accompli*.[34] However, the factory found that this approach led to legal challenges to environmental permits that then slowed construction.[35]

---

[27] Rooney et al. 2014, p 215.

[28] Stilgoe et al. 2006, p 19; Raman and Mohr 2014, p 12.

[29] Fiorino 1990.

[30] Gwen Ottinger 2013.

[31] Gwen Ottinger 2013; National Academies of Sciences, Engineering, and Medicine 2016, p 134.

[32] Fiorino 1990, p 228.

[33] Fiorino 1990, p 227. National Academies of Sciences, Engineering, and Medicine 2016, p 133.

Consequently, the factories reversed the order of their public engagement. Before developing the construction plan and hiring engineers, the factories engaged with the local community to understand their concerns.[36] Often, these concerns had little connection to environmental or health risk but rather pertained to odor or aesthetics, such as visible steam emissions.[37] After listening to these concerns, the factories instructed the engineers to design the plant to take these concerns into account, by for example, eliminating steam emissions.[38] The factory then submitted these designs to the appropriate regulators for legal approval.[39] When the permits were granted, the community had already agreed. The number of legal challenges fell dramatically and projects were completed more expeditiously.[40]

The social license concept has several advantages for proponents, the public at large and regulators. As the pulp mill example illustrates, focusing on consent facilitates expeditious pursuit by lessening opposition undertaken through legal action, legislation, civil protest or other means. For communities and stakeholders, a focus on consent increases the likelihood that proponents will meaningfully address community concerns rather than expert concerns, even in the absence of government regulation pertaining to those concerns.[41] Thus, the pulp mill above focused on aesthetics, a concern that regulators or engineers may have ignored otherwise. For regulators, considering a social license in the construction of a regulatory regime ensures that such regime is congruent with the requirements of a social license, and thereby helps project proponents achieve a social license. For all parties, a social license approach can lead to higher quality decisions,[42] almost as a form of peer review.

---

[34] Gunningham et al. 2004, p 327.

[35] *Id*.

[36] *Id*.

[37] Gunningham et al. 2004, pp 318–319.

[38] Gunningham et al. 2004, p 327.

[39] *Id*.

[40] *Id*.

[41] For nuclear technologies, which are heavily regulated, government regulation with regards to health and safety likely addresses health and safety concerns. However, there may be other values, such as the cultural significance of a power plant location, that government regulations is less able to address. Note that some commentators argue that the social license concept undermines the rule of law by giving local communities veto authority over otherwise approved projects. Canadian Association of Petroleum Landmen 2017. In some respects, the tension between local veto and legal authority is analogous to the tension between minority and majority rights in a democracy: minority groups have some inalienable rights, regardless of majority decision by vote.

Although a social license is not a binding legal obligation, legally enforceable obligations are often a key component of a social license.[43] In fact, in many, if not most, instances, social expectations exceed legal requirements.[44] The fact that a social license can often exceed legal obligations has been used to explain otherwise disparate behavior. For example, the social license concept can explain: (i) why some business enterprises undertake pollution control measures beyond their legal obligations;[45] (ii) why local communities have been able to block mining and energy projects that otherwise were approved through a formal legal process;[46] and (ii) why medical record data mining for public health purposes was terminated in the U.K.[47]

Note that the social license concept is not a panacea to the challenges of building new nuclear facilities. Project proponents adopting a social license approach are not guaranteed that the project will succeed. In fact, quite the opposite. Project proponents who undertake a social license approach must be genuinely open to the possibility that society, however defined for the particular project, will ultimately reject the project.[48]

A social license approach alone also cannot overcome nuclear power's contentious history. Prior public opposition to nuclear projects, especially nuclear waste repositories, can be viewed as a "profound breakdown of trust in the scientific, governmental, and industrial managers of nuclear technologies."[49] Psychological research validates the common-sense view that trust, once lost, is slowly, if ever, regained.[50] Thus, it would be naïve to presume that by simply adopting a social license approach, nuclear technologists can overcome decades of opposition. The entrance of new proponents with new technologies and in the context of climate change may provide an opening to rebuild trust with the public through a social license approach that takes the establishment and maintenance of trust seriously. However, it is also possible that even with a social license approach, trust cannot be rebuilt at this time.[51]

---

[42] Reed 2008, p 2419.

[43] Gunningham et al. 2004, p 329.

[44] Gunningham et al. 2004, pp 308, 329.

[45] Gunningham et al. 2004, p 308.

[46] Canadian Association of Petroleum Landmen 2017.

[47] Carter et al. 2015, p 2.

[48] Effective engagement requires that outcomes are "necessarily uncertain." Reed 2008, p 2426.

[49] Slovic et al. 1991, p 1606.

[50] Slovic et al. 1991, p 1606.

A social license approach also cannot overcome the economic challenges facing nuclear technologies, which are substantial. For example, in the U.S., existing nuclear plants are in many cases uneconomical to operate and many are either closing or seeking state subsidy to continue operating; only two new plants are currently under construction.[52]

### #.2.2 Society and Indicators of Consent

A social license is nebulous in part because both "society" and "consent" are nebulous themselves and context dependent. For any given project, the definition of "society" and "consent" depend on the nature of the project and the cultural expectations of whatever defines "society" for that project. In the environmental management literature, "society" for purposes of engagement, is often defined as "stakeholders," typically defined as "individuals who either affect or are affected by a decision or action."[53] In some circumstances, stakeholders can include non-human or non-living entities and future generations.[54] There is a field of inquiry, "stakeholder analysis," devoted to developing methods for identifying and classifying stakeholders.[55] Typically, stakeholders are identified before an engagement is undertaken, but in some cases, stakeholders are identified through an engagement process itself.[56] For some types of activities, the definition of stakeholders is a simple geographic boundary,[57] Jurisprudence regarding who can bring suit in court may be another source of inspiration for determining "society" for purposes of a social license.[58]

---

[51] Slovic et al. 1991, p 1606.

[52] Plumer 2018.

[53] Reed et al. 2009, p 1934.

[54] Reed 2008, p 2423.

[55] Reed et al. 2009.

[56] Reed 2008, p 2423.

[57] For example, the shoreline of an island hosting a genetically engineered mosquito trial, where the mosquito is not expected to leave the island, defined the boundary for societal consent via referendum. http://keysmosquito.org/oxitec-ox513a-trial/#148520330201-22f823b0-92bd.

[58] For example, in U.S. federal court, litigants must demonstrate that (i) they have "suffered a concrete and particularized injury that is either actual or imminent"; (ii) "the injury is fairly traceable to the defendant"; and (iii) "a favorable decision will redress the injury." *Mass. v. EPA* 549 U.S. 497, 517 (2007), *citing Lujan v. Defenders of*

For nuclear activities, "society" or "stakeholders" are hard to define due to the ability of radioactive materials to travel long distances and the fact that long-lived radioisotopes have the potential to affect future generations. Thus, "stakeholders" for a nuclear project likely includes far more than just individuals who live or work in the vicinity. Individuals located hundreds of miles away from a fission-based nuclear plant may see themselves as stakeholders given the risk of a catastrophic accident spreading radioactive contamination. Relatedly, individuals along nuclear material shipping routes may also see themselves as stakeholders in a decision to build nuclear facilities. Thus, stakeholders for nuclear facilities depend on weather patterns, local hydrology, and shipping routes and likely can only be identified during engagement.

Defining stakeholders for nuclear facilities is further complicated by the fact that under normal operation nuclear facilities do not present a significant risk to people or the environment. It is only during accidents, which are rare, that nuclear facilities have the potential to cause harm. A conservative approach is to include as a stakeholder any individual at risk of being impacted by a catastrophic accident, no matter how small the risk. A similar approach has been adopted by the International Law Commission with regards to transboundary harm in the environmental context,[59] and by the Implementation Committee of the Espoo Convention, which has argued that notification to parties to the Convention is required "unless a risk of significant adverse transboundary impact can be excluded."[60] As a practical example of this approach, the Paks nuclear power facility in Hungary, when considering construction of two new nuclear reactors, notified and sought comment from citizens from all seven neighboring countries, Switzerland and all European member states.[61]

How the stakeholders view consent is a further question, which depends on the expectations of the stakeholders themselves and, unfortunately, cannot be answered in general.

---

*Wildlife*, 504 U.S. 555, 560-561 (1992). In a possible analogy to nuclear risks, the U.S. Supreme Court has held that future injuries to the state of Massachusetts caused by greenhouse gasses meet this three-part test. *Id*. at 521.

[59] The International Law Commission has opined that risks of "causing significant transboundary harm" that trigger notification duties include "risks taking the form of a high probability of causing significant transboundary harm" and risks taking the form of "a low probability of causing disastrous transboundary harm." International Law Commission 2001, pp 5–6.

[60] Convention on Environmental Impact Assessment in a Transboundary Context Implementation Committee 2011 Citing decision IV/2 annex I, para. 54.

[61] United Nations Economic Commission for Europe 2017, p 34.

While consent could be a vote of elected officials, or even a community-wide referendum,[62] in some communities, it is conceivable that consent could be indicated by the assent of community elders. Alternatively, consent could be indicated by "the absence of widespread disagreement."[63]

### #.2.3 Key Principles

Because a social license focuses on public concerns rather than an expert assessment of risk, the exact procedures by which a social license is acquired and the concerns which it addresses depends on the activity seeking a social license. Engagement procedures that are appropriate for one type of activity or one type of community may be inappropriate or even counter-productive for a different activity or different community.[64] Furthermore, even for classes of activities for which best practice procedures have been established, the manner in which such procedures are followed are critical for success.[65] Thus, a comprehensive discussion of the procedures that should be undertaken to acquire a social license cannot be taken out of the context of the activity that a proponent is undertaking. Nevertheless, an analysis of prior projects that have successfully achieved a social license suggest that there are four generic principles: (i) engendering trust; (ii) transparency; (iii) meaningful public engagement; and (iv) protecting health, safety and the environment.

### #.2.3.1 Engendering Trust

A social license requires that proponents of an activity engender trust.[66] The requirement pervades all other principles and informs how project processes should be implemented. Proponents must convince stakeholders that (i) "they can be trusted to do the things they say they will do;" (ii) that they are honest with respect to the risks, harms and benefits of a proposed activity; (iii) that they are honest with how they intend to address these risks, harms and benefits; and (iv) that they pursue their actions in good faith, i.e., that they do not have a hidden agenda. Crucially, trust, once lost, is hard, if not impossible, to regain.[67]

---

[62] A trial of a genetically modified mosquito in the Florida Keys exemplifies both techniques. The local mosquito control board approved the trial and then put the trial to a local non-binding referendum. Citizens on one island approved the trial, while citizens on another island declined. The sponsoring private company and local mosquito control board then followed the results of the referendum. See http://keysmosquito.org/oxitec-ox513a-trial/#1485203300201-22f823b0-92bd.

[63] Bickerstaffe and Pearce 1980, p 330.

[64] Reed 2008, p 2424.

[65] Reed 2008, p 2425.

[66] Rooney et al. 2014, p 210; Hall et al. 2015, p 306. "Although there is no universally accepted definition of trust" trust is described as "having three main aspects: confidence, the belief that an individual or entity has the ability to do what they say they will do; integrity, the belief that an individual or entity is fair and just; and dependability, the belief that an individual or entity will do what they say they will do. Trust also depends on the available information that serves as the basis for judging these characteristics." National Academies of Sciences, Engineering, and Medicine 2016, p 136. *Citing* Hon and Grunig 1999.

*#.2.3.2 Transparency*

A social license further requires that project proponents are transparent. Transparency is universally advised in the social license literature and supports the need to engender trust.[68] Transparency to some degree is also often a legal requirement for nuclear power plant permitting.[69] Full transparency includes the sharing of all relevant information, including, *inter alia*, motivations, conflicts of interest, risks, and benefits.[70] Transparency provides the information by which stakeholders can evaluate (i) whether proponents have the ability to "do what they say they will do"; (ii) whether proponents will in fact "do the things they say they will [do]"; (iii) whether proponents are "fair and just;" (iv) whether proponents are motivated by their publically stated reasons or some other, perhaps improper, justification; and (v) whether proponents have properly addressed social values and concerns.[71]

However, full transparency can be difficult for activities that implicate national security or trade secrets. Project proponents may have a habit of withholding information.[72] Further, regulations that facilitate or require public engagement may presume that some information is confidential. In some cases, transparency may require a waiver from regulators.[73]

---

[67] Slovic et al. 1991, p 1606.

[68] Coglianese et al. 2008, p 927; Long and Scott 2013, p 49. ("To engender trust, the people or groups conducting or managing research should explain clearly what they are trying to accomplish, what they know and do not know, and the quality of the information [that] they have. They should reveal intentions, point out vested interests, and admit mistakes, and do all of this in a way that is frank and understandable[.]").

[69] For example, the U.S. Nuclear Regulatory Commission requires that all documents and correspondence relating to an application for a nuclear power plant construction permit and operating or combined license be made available to the public. U.S. Nuclear Regulatory Commission 2004, p 1.

[70] According to one mining executive, a social license is best acquire by "show[ing] anyone anything." Gunningham et al. 2004, p 327.

[71] Rooney et al. 2014, p 210; Hall et al. 2015, p 306; National Academies of Sciences, Engineering, and Medicine 2016, p 136. Coglianese et al. 2008, p 927.

[72] For example, the nuclear industry is often accused of having a habit of secrecy, stemming in part from its historical connection to weapon development and fear of public concern. See Bickerstaffe p 326-327.

[73] For example, in a genetically engineered mosquito trial, the sponsoring company gave permission for the FDA to open its draft environmental assessment to public comment in order to facilitate effective public engagement. "Letter from Oxitec Ltd. To FDA DDM re: Draft Environmental Assessment for Investigational Use of Aedes

Note that the manner and form of information release determines the extent to which transparency supports a social license. Release of massive amounts of highly technical information will be likely perceived as an attempt to bury unflattering information under the cover of "transparency." Information must be released in a form that is easy to acquire, clearly articulated and easy to understand.[74] For highly technical projects, such as nuclear, proponents may need to provide education[75] or employ skilled facilitators who can "translate" technical information so that non-experts can effectively understand and participate.

### #.2.3.3 Meaningful Public Engagement

Public engagement is universally advised in the social license literature and is the essential feature of the social license approach.[76] It is essential for engendering trust. Engagement is frequently discussed in the context of developing and deploying new technologies[77] Engagement of some kind is also often a legal requirement of environmental,[78] administrative[79] and international law.[80] In the U.S., nuclear power plant permitting by the U.S. Nuclear Regulatory Commission requires multiple public hearings and opportunities to comment, both orally and in writing, on site selection permits, environmental reviews, power plant design, construction permits, operating licenses, and combined licenses.[81]

---

aegypti OX513A, available at https://www.regulations.gov/document?D=FDA-2014-N-2235-1294. The FDA is statutorily required to maintain the confidentiality of information submitted by research proponents, absent such a waiver. 21 CFR §§ 20.61(c), 25.50(b), 514.11.

[74] Coglianese et al. 2008, p 926.

[75] Reed 2008, p 2422.

[76] Coglianese et al. 2008.

[77] National Academies of Sciences, Engineering, and Medicine 2016, p 131. When discussing the benefits of a public process of review of gene transfer experiments, a National Academy report remarked that "By engaging the public in a focused discussion on the technology and its potential societal impacts, the RAC engendered trust and credibility." Institute of Medicine 2014, p 5.

[78] In the U.S, federal agencies undertaking major actions that significantly affect the quality of the human environment must complete an environmental impact statement and must solicit comments from the public as part of competing such statement. 40 CFR §§ 1502.3, 1503.1(a)(4) (2018) and 42 U.S.C. § 4332 (C).

[79] In the U.S., federal agencies are required to provide an opportunity for "interested persons" to participate in agency rulemaking proceedings through written comments and other means. 5 USC § 553(c) (2018).

[80] The Espoo Convention requires transboundary public engagement for projects that are "likely to cause a significant adverse transboundary impact." *Convention on Environmental Impact Assessment in a Transboundary Context*, Article 3.8.

[81] U.S. Nuclear Regulatory Commission 2004.

Although engagement is listed here third, it is recommended to be undertaken as early as possible, in order for proponents to understand societal concerns and to give proponents an early opportunity to shape their subsequent procedures to meet these concerns.[82] Nonetheless, the effectiveness of public engagement in facilitating the acquisition of a social license depends critically on the manner in which public engagement is pursued.[83] Ineffective public engagement can be counter-productive and breed public cynicism if it creates the impression that proponents are not genuinely interested in pursuing engagement or have something to hide.

Effective public engagement is more than simply accepting public comments or sharing intended plans or results. It has been described as "seeking and facilitating the sharing and exchange of knowledge, perspectives, and preferences between or among groups who often have differences in expertise, power, and values."[84] Engagement cannot be a one-way lecture where proponents simply explain what they intend to do and why the public should accept the activity. Engagement in which members of the public are simply "educated" regarding a choice that has already been made, and asked to comment or vote, is often counterproductive as it instills a sense of condescension, or "paternalism of expertise."[85] Engagement must be a two-way conversation in which proponents learn from the public. A two-way conversation can be challenging for technical projects, which have a tendency to use public engagement to let "people see the experts at work."[86] To counter this tendency, proponents should engage in meaningful discourse, dialog, and negotiation with stakeholders in a genuine process[87] that (i) respects, acknowledges and takes into account divergent views and values;[88] (ii) is inclusive and minimizes power differentials between stakeholder groups;[89] and (iii) allows stakeholders to participate, to some degree, in the proponent's decision making process.[90] Engagement is also an ongoing process that is not limited to one time events; rather it is an iterative process that requires "attention to multiple types of communication, deliberation, relationship building, reflection, and empowerment."[91]

---

[82] Reed 2008, p 2422.

[83] "The outcome of any participatory process is far more sensitive to the manner in which it is conducted than the tools that are used." Reed 2008, p 2425.

[84] National Academies of Sciences, Engineering, and Medicine 2016, p 131.

[85] Bickerstaffe and Pearce 1980, p 323.

[86] Stilgoe et al. 2006, p 19; Raman and Mohr 2014, p 12.

[87] Rooney et al. 2014, p 214.

[88] Rooney et al. 2014, p 215.

[89] Reed 2008, p 2422; Rooney et al. 2014, p 214.

[90] Coglianese et al. 2008, p 926.

Crucially, engagement must be *meaningful*. Proponents must offer a genuine opportunity to affect the course of a project. Stakeholders must be sufficiently informed that they can offer substantive suggestions.[92] Meaningful engagement is achieved through a process by which all interested stakeholders can participate and have their interests heard even-handedly.[93] The process must demonstrate that proponents have undertaken "a good-faith effort toward respectful listening, creative compromise, and flexible practice."[94] Skilled facilitators can be useful to help translate technical information and mediate power and knowledge differentials between stakeholders and proponents.[95]

There are compelling practical reasons for proponents to pursue engagement. *First*, proponents cannot learn what concerns are elicited by a proposed project without engagement. These concerns are often not self-evident. Without public engagement, proponents can only guess what concerns the public has, or worse, only address proponent's concerns. The pulp mill example above, in which the community was most concerned with the aesthetic impact instead of the health and safety impacts, dramatically illustrates this point. *Second*, engagement improves outcomes. It provides a means of obtaining additional information[96] and leads to better and more informed decisions due to the fact that non-experts, in some circumstances, are better able to judge risk than experts.[97] Better decisions increases the likelihood that an activity will not harm health or the environment and will respect society's values. In turn, better and more informed outcomes create a positive feedback-loop that further engenders trust for future engagement.[98] In effect, engagement acts as a form of peer review.

---

[91] *Id.*

[92] According to one engagement expert, "It is not enough simply to provide stakeholders with the opportunity to participate in decision-making […]. [T]hey must actually be able to participate." Reed 2008, p 2422.

[93] Coglianese et al. 2008, p 927.

[94] National Academies of Sciences, Engineering, and Medicine 2016, p 134.

[95] Reed 2008, pp 2422, 2425.

[96] Coglianese et al. 2008, p 927; National Academies of Sciences, Engineering, and Medicine 2016, p 133.

[97] Fiorino 1990, p 227. Non-experts "see problems, issues, and solutions that experts miss." They have a sensitivity "to social and political values that experts' models [do] not acknowledge." *Id.* Further, they may have a better capacity for "accommodating uncertainty and correcting errors over time through deliberation and debate." *Id.*

*#.2.3.4 Protecting Health, Safety and the Environment*

The entity that seeks a social license must assess health, safety and environmental risks of a proposed activity, and either take steps to minimize such risk or abandon projects that pose too great of a risk.[99] For nuclear projects, this principle often draws the most attention, debate and litigation.[100] However, from a social license perspective, it is the least controversial as protection of health, safety and the environment is presumed.

#.2.4 The Role of Regulation

Regulation has a strong role to play in facilitating the acquisition of a social license. Regulation can make explicit the means by which proponents address societal concerns and expectations, thereby freeing project proponents from conducting their own evaluation. Regulation can also assign responsibilities for adhering to different principles of a social license. For example, regulation can require government agencies to undertake engagement as part of issuing a legal permit or license. Regulation can standardize procedures for classes of similar types of projects, such human subject research. Perhaps most critically, regulation can engender trust by enforcing, through civil or criminal penalty, transparency, public engagement, and health, safety and environmental protection.

On the other hand, regulation can also impair a social license. For example, regulation or legislation that specifies a controversial end result, such as the location of a nuclear waste repository, is likely to be counterproductive as such specificity impairs genuine public engagement and may lead stakeholders to undertake legal or other action outside of the engagement process.[101]

---

[98] National Academies of Sciences, Engineering, and Medicine 2016, p 135.

[99] Gunningham et al. 2004, p 314; Rooney et al. 2014, p 210.

[100] For example, nuclear power advocates occasionally argue that existing health, safety and environmental regulations are excessively expensive and stifle innovation. See Institute for Energy Research 2018.

[101] See the nuclear waste repository case studies below.

Further, it should be emphasized that regulatory compliance alone is not sufficient in itself for an activity to achieve a social license.[102] There is a risk, especially for heavily regulated activities like nuclear, that proponents and governments become lulled into a false sense of security with regard to a social license through regulatory compliance. Regulation can be inflexible, leading to an out-of-date regulatory regime that is unable to meet society's expectations. New types of societal concerns or new safety risks may arise that were not contemplated when regulations were created. Regulation may not include sufficient transparency, or the public engagement specified by regulation may not meet the unique circumstances of the activity or the community hosting the project. Moreover, regulation may not provide a means of assessing society's "consent."

There are many examples of projects that undertake measures beyond regulatory compliance. In fact, a defining characteristic of the social license approach is to ask what else should be done. Thus, the pulp mill mentioned above undertook engagement prior to applying for a permit and addressed concerns, such as aesthetics, that are not related to health or safety regulations. As another example, a genetically modified mosquito trial waived its right to a confidential environmental impact statement in order to facilitate meaningful public comment,[103] and sought approval from residents of the island hosting the trial through a vote of public officials and a non-binding referendum,[104] neither of which were required under federal regulations for genetically engineered organisms.[105]

#.2.5 The Role of Project Proponents and Governments

A critical question for the social license approach is which entity or entities takes responsibility for addressing society's concerns and achieving society's consent. In particular, who has responsibility for: (i) evaluating whether compliance with existing regulations is sufficient for a social license or whether additional actions or procedures are necessary; (ii) undertaking engagement and disclosing relevant materials; and (iii) ultimately determining whether consent has been achieved. Typically, these responsibilities are divided between project proponents, i.e., the entity that will actually undertake the project, and government agencies, which either regulate or provide financial funding. Note that there are also cases in which the government itself is the project proponent.

---

[102] Gunningham et al. 2004, pp 308, 329.

[103] *Supra* note 73.

[104] *Supra* note 62.

[105] For this particular trial, the U.S. Food and Drug Administration exercised regulatory oversight under its authority over animal drugs. U.S. FDA, Center for Veterinary Medicine 2017, p 6. FDA approval was conditioned on an environmental assessment and a determination that the investigation was neither unsafe nor "otherwise contrary to the public interest." *Id*. FDA regulations do not discuss local community acceptance.

For heavily regulated or particularly risky activities, such as nuclear technologies, there is perhaps a tendency for a governmental actor to take primary responsibility for all aspects of a social license, either by crafting appropriate regulation or by direct action, including promulgating information and undertaking public engagement. A strong governmental role has at least four advantages. *First*, a governmental role frees proponents to focus on their particular expertise. This advantage is particularly pertinent for classes of projects that would be overwhelmed by achieving a social license, such as academic research. *Second*, governments have a unique ability to garner resources and advice to evaluate risks, some of which, such as national security, only governments can truly evaluate. *Third*, governments have a unique ability and resources to undertake public engagement. *Fourth*, governments, through their enforcement power, can engender trust, especial with regards to health, safety and environmental protection.

On the other hand, there are at least three disadvantages for placing responsibility on governments. *First*, as mentioned before, governments can be less agile than non-governmental actors and slower at responding to changing social expectations, either because of bureaucratic inertia or legal constraints. *Second*, national government actors or regulation may not be sensitive to local concerns. *Third*, government involvement can exacerbate power differences between proponents and society, to the detriment of meaningful and effective engagement.

Often, responsibilities for a social license are shared between project proponents and governments. For large, novel, unique, or controversial projects, for which social concerns and effective engagement procedures are uncertain at the outset of the project, the proponent often takes prime responsibility for securing the social license, supported by a governmental role to ensure health, safety and environmental protection. For example, a genetically engineered mosquito trial went beyond regulatory requirements for transparency, engagement and consent.[106] On the other hand, for routine and repeated activities, the government often takes the lead by creating regulations or codes of conduct. For example, both publicly and privately funded human subject research in the U.S. is governed by regulations and codes-of-conduct requiring local Institutional Review Board approval.[107] A social license for human subject research is achieved, in part, through public engagement in the creation and periodic revision of these regulations[108] and though the operation of Institutional Review Boards.

---

[106] *Supra* note 62, 73.

## #.3 Nuclear Case Studies

The history of nuclear waste repositories illustrates the social license concept, especially as it pertains to nuclear technologies.

### #.3.1 Yucca Mountain Waste Repository

The story of the Yucca Mountain nuclear waste repository in the U.S. state of Nevada illustrates how neglecting a social license approach can imperil a project. Since 1983, the U.S. Department of Energy ("U.S. DOE") has spent more than $13 billion evaluating the safety of Yucca Mountain.[109] Nevertheless, the repository remains in limbo. In 2010, the U.S. DOE withdrew its application to the U.S. Nuclear Regulatory Commission; however, a recent budget proposal has revived the repository's prospects.[110]

Explanations for Yucca Mountain's failure have been widely discussed and include the challenges of the geology,[111] the manner by which the DOE engaged, or did not engage, with the Nevada public, and the lack of consent by the State of Nevada to the site selection process.[112] Another explanation is that the U.S. Congress made it difficult, if not impossible, for the U.S.

---

[107] See 82 FR 7149 (January 19, 2017) and 45 CFR §§ 46.101, 122, 123. Strictly speaking, U.S. federal regulations on human subject research only pertain to those projects funded by the U.S. federal government. However, most institutions adopt these regulations as their own rules, regardless of funding source. See for example, https://cuhs.harvard.edu/procedures/institutional-authority. In addition, the U.S. FDA requires studies submitted in support of a new drug or medical device to have been conducted in conformance with federal human subject research regulations. 21 CFR 56.103 (2018).

[108] Human subject research regulations in the U.S. were recently revised in 2017 using a public engagement process. 82 FR 7149 (January 19, 2017). Before 2018, they were most recently revised in 2005. *Id.*

[109] Katherine Ling, NY Times, "Budget Will Eliminate Yucca Nuclear Waste Repository Says Sen. Reid." February 1, 2010. http://www.nytimes.com/gwire/2010/02/01/01greenwire-budget-will-eliminate-yucca-nuclear-waste-repos-9897.html; Ewing and Hippel 2009.

[110] Zhang 2017.

[111] Yucca Mountain is pervaded by geologic fractures that could transport radioactive materials to the local aquifer. Murphy 2006, p 49. There is a 5% chance that the radiation dose by contaminates in the aquifer will double an individual's annual radiation exposure. Whipple 2006, p 61.

[112] Blue Ribbon Commission on America's Nuclear Future 2012, p 23.; Ewing and Hippel 2009, p 151.

DOE to secure a social license. Under the 1987 Amendment to the Nuclear Waste Policy Act ("NWPA"), Congress designated Yucca Mountain as the sole location for a nuclear waste repository.[113] The U.S. Department of Energy ("US DOE") was tasked with evaluating the health and safety of such storage, subject to confirmation by the U.S. Nuclear Regulatory Commission ("NRC") and Environmental Protection Agency ("EPA").[114] If these agencies agreed, the US DOE was to construct and operate the repository. This siting process focused exclusively on health, safety and environmental protection and neglected the other principles required for a social license.[115] This focus was counter-productive. The emphasis on science blinded the US DOE leadership and the US Congress to the real issue facing the repository: whether Nevada was willing to host the repository, even if it was established to be safe. Residents and government officials in Nevada were never supportive of the repository. Opposition was so fierce, that at one point the state denied a water permit for bore-hole drilling and the DOE sued in federal court to enjoin Nevada from enforcing its own water laws.[116]

The siting process also precluded meaningful engagement. Because Congress legislated that the repository would be at Yucca Mountain as long as experts deemed it safe, there was no meaningful role for public input. Regardless of public comment, if the U.S. DOE, NRC and EPA determined that the repository would be safe, it would be built. This meant that public engagement that did took place[117] was merely an opportunity to let "people see the experts at work" rather than opening up "expertise to new questions and perspectives."[118]

#.3.2 Swedish & Finnish Waste Repositories

In contrast to the U.S., Sweden and Finland explicitly adopted a social license approach and have had a comparative success. The Finnish repository at Olkiluoto is now under construction[119] and the Swedish repository at Forsmark is pending licensing approval.[120] Although both nations began their site selection process by focusing on technical issues, they realized that local consent was as important, if not more important, than technical viability.[121]

---

[113] Cotton 2006, p 36.; Nuclear Waste Policy Act of 1982 and as amended 1987, 42 U.S.C. § 10101 *et seq*. (2017).

[114] *Id*.

[115] Dawn Stover, "The "scientization" of Yucca Mountain" 12 October, 2011. Bulletin of the Atomic Scientists. Available at https://thebulletin.org/scientization-yucca-mountain.

[116] See *U.S. vs. Nevada*, No. 2:00-CV-0268-RLH-LRL, ORDER on Second Motion for Preliminary Injunction-#120 (D. Nev. August 31, 2007), *available at* http://www.energy.ca.gov/nuclear/yucca/documents/AG-155-2007-000513.pdf.

[117] The DOE undertake numerous public engagement events. For example, between May and August of 2001 alone the DOE held 66 public hearings. U.S. Senate Committee on Environment and Public Works Majority Staff 2006.

[118] Raman and Mohr 2014, pp 10, 12.

In Sweden, the nuclear waste company, SKB, sought community consent for preliminary study from all 286 Swedish municipalities; three agreed.[122] SKB also performed feasibility studies in five municipalities that were already hosting a nuclear facility.[123] In 2000, eight sites in total were identified as technically and socially suitable for further investigation; two of these sites were selected and consented to in-further study.[124] In 2009, SKB selected Forsmark due to superior geology.[125] In 2011, SKB submitted the application for construction to the Swedish Radiation Safety Authority and the Environmental Court.[126]

In Finland, the nuclear waste company, Posiva, took a slightly different approach, but with the same result. After extensive geological surveys of Finland, Posiva selected four locations for further study.[127] A full environmental impact assessment was conducted for each location.[128] Crucially, each location and surrounding communities were given a veto over a final siting decision.[129] At the conclusion of the environmental impact assessment and a public engagement process, two communities expressed the highest interest in hosting a repository.[130] Of these, Olkiluoto was proposed in 1999 because of the larger geographic area available and the fact that more spent fuel was already located near the proposed site.[131] The location was ratified by the Finish parliament in May 2001 and a construction license was granted in 2015.[132]

---

[119] Fountain 2017.

[120] http://www.skb.com/news/a-week-in-osthammar-with-focus-on-the-environment/

[121] In 1985, the Swedish nuclear waste company, SKB, drilled exploratory bore-holes at 10 locations throughout Sweden. Lidskog and Sundqvist 2004. Local opposition to the drilling, however, led SKB to change its strategy: instead of focusing on the best site at the start of the process, the company decided to focus on many suitable sites that were all socially acceptable. *Id*. at 261. As explained by SKB: "The site where the final repository is built must fulfill two fundamental requirements: There must be bedrock that permits long-term safe disposal, and there must be political and popular support in the concerned municipality and among nearby residents." Swedish Nuclear Fuel and Waste Management Company 2011, p 19. According to Timo Aikas, a former executive of the Finish nuclear waste company, Posiva, initially, "[Posiva] ran into difficulties because we tried to behave as industry did back then – we'd decide and announce. […] Very soon we learned that we had to be very open. […] This openness and transparency creates trust." Fountain 2017.

[122] Lidskog and Sundqvist 2004, p 261.

[123] *Id*. at 262.

[124] Swedish Nuclear Fuel and Waste Management Company 2011, pp 19–20.

[125] *Id*.

[126] http://www.skb.com/future-projects/the-spent-fuel-repository/our-applications/

[127] The final disposal facility for spent nuclear fuel, At 4-7. http://www.posiva.fi/files/738/The_final_disposal_facility_for_spent_nuclear_fuel_small.pdf.

[128] *Id*.

[129] Curry 2017.

## #.4 Application to Nuclear Technologies

### #.4.1 Nuclear Waste Siting

The nuclear waste repository case studies illustrate the importance of consent and meaningful engagement, both points that have been made before.[133] The social license discussion leads to five concrete recommendations to achieve a social license:

*First*, in the U.S., nuclear power proponents should not advocate for Yucca Mountain. Yucca Mountain may be a safe location for nuclear waste. However, the prior process has impaired trust to such an extent that societal consent will likely never be achieved under the current mandated process. Thus, any attempt to revive Yucca Mountain, absent a complete restart in the siting process, is likely to be counter-productive to nuclear waste disposal.

*Second*, a governmental agency should not take prime responsibility for a nuclear waste repository. A governmental agency should be responsible for protecting health, safety and the environment, and may have a role to play in facilitating engagement. But it should not be primarily responsible for the social license. Nuclear waste repositories are new, large, and unique. Although the key principles of a social license are well established, societal concerns and effective procedures are uncertain and influx. Thus, the project proponent needs to be nimble and learn from its experience as engagement proceeds – it cannot be constrained by a process that is created a priori through regulation or legislation. Both Sweden and Finland adjusted their siting process over the decades that they pursued their repositories. This flexibility was possible, in part, because both Swedish and Finish nuclear power utilities were responsible for siting their waste repositories rather than the Swedish or Finish governments.[134]

---

[130] http://www.posiva.fi/en/final_disposal/selecting_the_site_the_final_disposal_at_olkiluoto#.Wk6JGbenGDI.

[131] *Id*.

[132] http://www.posiva.fi/en/final_disposal/general_time_schedule_for_final_disposal#.Wk6VF7enGDI.

[133] Blue Ribbon Commission on America's Nuclear Future 2012.

[134] The Swedish and Finnish governments were heavily involved in assessing health and safety.

Page 22 of 29

Furthermore, given the long history of nuclear technology's connections to government weapons programs that had limited, if any, public input, a non-governmental proponent is likely to have more credibility with regards to meaningful engagement. The power difference between a non-governmental proponent and host community is far less than a national government. In this context, engagement between a non-governmental proponent and stakeholders is more likely to be perceived as a true conversation rather than a one-way lecture. The danger of power differences is apparent in the Yucca Mountain example. In a dispute over water rights during technical feasibility studies, the U.S. DOE claimed that the Nuclear Waste Policy Act pre-empted Nevada's water laws,[135] essentially sending a message that the DOE would use its authority to do whatever it saw fit, regardless of local concerns, undercutting public engagement and trust. A non-governmental entity does not have such authority to override a local community's own policies, and thus, would not be able to exercise such blatant power.

*Third*, proponents should only undertake studies in willing communities.[136] Consent at the start of the siting process will facilitate future engagement. Seeking consent before evaluation is also a good-faith gesture on the part of proponents that will engender trust.

*Fourth*, at least two sites should be evaluated to the same level of detail before a final site is selected. The in-depth technical evaluation of at least two sites in Finland and Sweden likely made selection of one location possible. Although multiple evaluations may appear inefficient, multiple sites signals to communities that they have a genuine choice with regards to the repository – other options are available if a community choses to not host a repository.

*Fifth*, communities should always have a veto over a siting decision.[137] Like "consent," a veto is nebulous and depends on the context of the community that is considering a repository. Elected officials, leaders, or a community vote may wield the veto. Nevertheless, a veto complements the advantages of multiple site investigations and further strengthens the meaningfulness of public engagement. For example, a veto means that if a proponent neglects public comments, a community can reject the project.

---

[135] *U.S. v. Morros*, 268 F.3d 695, 699 (9th Cir. 2001).

[136] However, a preliminary geologic screen of suitable communities likely does not require community consent to the extent that such screening requires limited, if any, sampling or drilling.

[137] In the event that no community agrees to host a waste repository, the siting process would need to begin again.

#.5.2 Advanced Fission and Fusion Power Plants

Many startups and academic groups are pursuing novel fission and fusion power plant designs. Although none of these designs are near operational, it is prudent for researchers and developers to begin thinking about how to secure a social license for future commercial power plants, if their research is successful, and thus, avoid some of the controversies that have long plagued nuclear technologies. Fission and fusion are very different technologies that pose very different waste, accident and non-proliferation risks. The precise means by which either technology secures a social license, and the regulatory mechanisms that are most appropriate for either technology's specific risks are likely very different. Nevertheless, the discussion above suggests the following four high-level recommendations that apply to either fission or fusion-based reactors for securing a social license: *First*, the advanced fission and fusion communities should embrace responsibility for achieving a social license and not rely on regulatory compliance alone. This means that they should (i) evaluate the extent to which existing laws and regulations are effective in facilitating a social license; (ii) think critically about the role of government in securing a social license; (iii) take actions themselves, in light of the role of government, to secure a social license; and (vi) propose regulatory reforms that complement their efforts and that would help facilitate a social license.

*Second*, because a social license requires a sensitivity to public concerns rather than expert concerns, the advanced fission and fusion communities should solicit and listen to concerns of the public. For example, the public may have more worry than experts with regards to waste disposal, radioactive material storage and transport, the risk of fires and explosions, whether caused by an accident or sabotage, and the risk of increasing nuclear proliferation. Regardless of how experts feel about these issues, the advanced fission and fusion communities should take them seriously through both engineering design and the development of regulation appropriate for these concerns.

*Third*, following the nuclear waste siting discussion, advanced nuclear plants, whether research, demonstration or commercial, should only be sited in communities that actively welcome these facilities. Forcing a community to host a new facility is likely to be counter-productive for either industry as a whole.

*Lastly*, fission and fusion proponents should embrace stewardship of the entire lifecycle of their technology, including waste disposal and plant decommissioning. A longstanding and principle concern for nuclear power plants has always been disposal of waste and decommissioning of plants. In order to secure a social license, advanced nuclear power technologies, whether fission or fusion-based, will need to directly address this concern. Advanced fission proponents will need a strategy in light of the challenge of disposing of spent fuel from existing fission plants, and in the U.S., the need to distance themselves from Yucca Mountain. Fusion will not be immune from this concern, even if the amount of waste is smaller and the waste itself radioactive for a far shorter duration than fission. If fusion becomes commercially viable, the fusion industry can address this concern head-on by including waste disposal and decommissioning in their business plans. For example, companies operating fusion plants could set-aside funds for disposal and decommissioning. Companies could start planning for decommissioning and disposal at the same time that plants are under design and construction.[138] Such plans could include, for example, subsidiaries or collaborations that start the siting process for their particular waste needs at the same time, or even before, commercial fusion construction.

# #.5 Conclusion

This Chapter has provided an overview of the concept of a social license, used the concept to explain why some nuclear waste repositories are under construction while others have failed, and detailed concrete recommendations for future nuclear waste repositories and advanced fission and fusion-based power plants to secure a social license. The analysis presented here suggests that addressing societal concerns, whatever they may be, through a meaningful and transparent engagement process that engenders trust, is as important as technical innovation in order for nuclear technologies have a role in responding to climate change.

Suggestions for meaningful public engagement are not new. However, experience in fields other than nuclear provides insight into how nuclear technologists can best secure a social license. The analysis presented in this Chapter leads to four high-level lessons for nuclear technologists: *First*, technologists cannot rely on government regulators to secure the social license on their behalf; technologists must take responsibility for securing the social license for themselves, with support from government regulation and government agencies. This means that proponents should be prepared to go beyond regulatory requirements for safety, waste disposal, engagement, transparency, and other stakeholder concerns.  *Second*, while being prepared to go beyond government regulation, technologists should also consider how government regulation can support a social license, not just by addressing health, safety and environmental concerns, but also by addressing other concerns, such as the management of nuclear waste, and advocate for such regulation.  *Third*, because the manner in which engagement is conducted is far more important than the tools used for engagement, technologists should craft the engagement process for any particular project to meet the unique circumstances of the project and the stakeholders involved.  Participating stakeholders should be given the knowledge and tools they need to make informed comments, facilitators should be used to overcome knowledge and power differences, and project proponents should be genuinely open and respond to stakeholder input.  *Lastly*, technologists should embrace responsibility for a social license for their technology's full life cycle.  A common critique of nuclear technologies is that it forces future generations to deal with waste generated today.  In order to secure a social license, technologists will need to directly address this concern.

---

[138] In some circumstances, it may even be possible to co-locate fusion plants and waste disposal facilities to minimize transport of radioactive material.

# References

Bankes N (2015) The social license to operate: mind the gap. In: University of Calgary Faculty of Law Blog on Developments in Alberta Law. http://ablawg.ca/wp-content/uploads/2015/06/Blog_NB_SLO_June2015.pdf. Accessed 23 Mar 2017

Bickerstaffe J, Pearce D (1980) Can There Be a Consensus on Nuclear Power? Social Studies of Science 10:309–344.

Blue Ribbon Commission on America's Nuclear Future (2012) Report to the Secretary of Energy. U.S. Department of Energy

Canadian Association of Petroleum Landmen (2017) Social License to Operate. http://landman.ca/2017/03/13/social-license-operate/. Accessed 23 Mar 2017

Carter P, Laurie GT, Dixon-Woods M (2015) The social licence for research: why *care.data* ran into trouble. Journal of Medical Ethics 41:404–409. doi: 10.1136/medethics-2014-102374

Coglianese C, Kilmartin H, Mendelson E (2008) Transparency and public participation in the federal rulemaking process: Recommendations for the new administration. Geo Wash L Rev 77:924.

Convention on Environmental Impact Assessment in a Transboundary Context Implementation Committee (2011) Opinions of the Implementation Committee (2001-2010).

Cotton T (2006) Nuclear Waste Story: Setting the Stage. In: Macfarlane A, Ewing RC (eds) Uncertainty Underground: Yucca Mountain and the Nation's High-level Nuclear Waste. MIT Press, Cambridge, MA,

Curry A (2017) What Lies Beneath. The Atlantic

Dixon-Woods M, Ashcroft RE (2008) Regulation and the social licence for medical research. Med Health Care and Philos 11:381–391. doi: 10.1007/s11019-008-9152-0

Edenhofer O (ed) (2014) Climate change 2014: mitigation of climate change: Working Group III contribution to the Fifth Assessment Report of the Intergovernmental Panel on Climate Change. Cambridge University Press, New York, NY

Erikson K (1990) Toxic Reckoning: Business Faces a New Kind of Fear. Harvard Business Review 68:118–126.

Ewing R, Hippel FV (2009) Nuclear Waste Management in the United States-Starting Over. Science 325:151–2. doi: 10.1126/science.1174594

Falk J (1982) Global fission - The battle over nuclear power.

Fiorino DJ (1990) Citizen Participation and Environmental Risk: A Survey of Institutional Mechanisms. Science, Technology, & Human Values 15:226–243.

Fountain H (2017) On Nuclear Waste, Finland Shows U.S. How It Can Be Done. The New York Times

Gunningham N, Kagan RA, Thornton D (2004) Social license and environmental protection: why businesses go beyond compliance. Law & Social Inquiry 29:307–341.

Gwen Ottinger (2013) Changing Knowledge, Local Knowledge, and Knowledge Gaps: STS Insights into Procedural Justice. Science, Technology, & Human Values 38:250–270. doi: 10.1177/0162243912469669

Hall N, Lacey J, Carr-Cornish S, Dowd A-M (2015) Social licence to operate: understanding how a concept has been translated into practice in energy industries. Journal of Cleaner Production 86:301–310. doi: 10.1016/j.jclepro.2014.08.020

Hon L, Grunig J (1999) Guidelines for MEasuring Relationships in Public Relations. Institute for Public Relations

Institute for Energy Research (2018) Regulations Hurt Economics of Nuclear Power. In: IER. https://instituteforenergyresearch.org/analysis/regulations-hurt-economics-nuclear-power/. Accessed 3 May 2018

Institute of Medicine (2014) Oversight and Review of Clinical Gene Transfer Protocols: Assessing the Role of the Recombinant DNA Advisory Committee. National Academies Press, Washington, D.C.

International Atomic Energy Agency (ed) (2008) Chernobyl: looking back to go forward: proceedings of an International Conference on Chernobyl: Looking Back to Go Forward ... held in Vienna, 6-7 September 2005. International Atomic Energy Agency, Vienna, Austria

International Energy Agency (2015) Special Report on Climate Change. OECD Publishing, Paris

International Law Commission (2001) Draft Articles on Prevention of Transboundary Harm from Hazardous Activities.

Lassiter J (2018) Op-Ed: Why Private Investors Must Fund "New Nuclear" Power Right Now. In: HBS Working Knowledge. http://hbswk.hbs.edu/item/op-ed-why-private-investors-must-fund-new-nuclear-power-right-now. Accessed 25 Apr 2018

Lidskog R, Sundqvist G (2004) On the right track? Technology, geology and society in Swedish nuclear waste management. Journal of Risk Research 7:251–268. doi: 10.1080/1366987042000171924

Long JC, Scott D (2013) Vested Interests and Geoengineering Research. Issues in Science and Technology 29:45–52.

Macfarlane A, Ewing RC (eds) (2006) Uncertainty Underground: Yucca Mountain and the Nation's High-level Nuclear Waste. MIT Press, Cambridge, MA

Mufson M (1982) Psychosocial Aspects of Nuclear Power: A Review of the International Literature. In: American Psychiatric Association (ed) Psychosocial aspects of nuclear developments. American Psychiatric Association, Washington, D.C,

Murphy W (2006) Regulating the Geologic Disposal of High-Level Nuclear Waste at Yucca Mountain. In: Macfarlane A, Ewing RC (eds) Uncertainty Underground: Yucca Mountain and the Nation's High-level Nuclear Waste. MIT Press, Cambridge, MA,

National Academies of Sciences, Engineering, and Medicine (2016) Gene Drives on the Horizon: Advancing Science, Navigating Uncertainty, and Aligning Research with Public Values. National Academies Press, Washington, D.C.

National Science Advisory Board for Biosecurity (2016) Recommendations for the Evaluation and Oversight of Proposed Gain-Of-Function Research. National Science Advisory Board for Biosecurity

Nuclear Energy Institute (2018) Ensuring the Future of U.S. Nuclear Energy Creating a Streamlined and Predictable Licensing Pathway to Deployment.

Otway HJ, Maurer D, Thomas K (1978) Nuclear power: The question of public acceptance. Futures 10:109–118. doi: 10.1016/0016-3287(78)90065-4

Plumer B (2018) The U.S. Backs Off Nuclear Power. Georgia Wants to Keep Building Reactors. The New York Times

Raman S, Mohr A (2014) A social licence for science: capturing the public or co-constructing research? Social Epistemology 28:258–276.

Reed MS (2008) Stakeholder participation for environmental management: A literature review. Biological Conservation 141:2417–2431. doi: 10.1016/j.biocon.2008.07.014

Reed MS, Graves A, Dandy N, et al (2009) Who's in and why? A typology of stakeholder analysis methods for natural resource management. Journal of Environmental Management 90:1933–1949. doi: 10.1016/j.jenvman.2009.01.001

Reed MS, Vella S, Challies E, et al (2018) A theory of participation: what makes stakeholder and public engagement in environmental management work? Restoration Ecology 26:S7–S17. doi: 10.1111/rec.12541

Rooney D, Leach J, Ashworth P (2014) Doing the Social in Social License. Social Epistemology 28:209–218. doi: 10.1080/02691728.2014.922644

Slovic P (1987) Perception of risk. Science 236:280–285. doi: 10.1126/science.3563507

Slovic P (1996) Perception of Risk from Radiation. Radiation Protection Dosimetry 68:165–180. doi: 10.1093/oxfordjournals.rpd.a031860

Slovic P, Flynn JH, Layman M (1991) Perceived Risk, Trust, and the Politics of Nuclear Waste. Science 254:1603–1607. doi: 10.1126/science.254.5038.1603

Sorbom BN, Ball J, Palmer TR, et al (2015) ARC: A compact, high-field, fusion nuclear science facility and demonstration power plant with demountable magnets. Fusion Engineering and Design 100:378–405. doi: 10.1016/j.fusengdes.2015.07.008

Stilgoe J, Irwin A, Jones K, Demos (Organization : London E (2006) The received wisdom: opening up expert advice. Demos, London

Swedish Nuclear Fuel and Waste Management Company (2011) Application for license under the nuclear activities act.

The Economist (2013) Limiting the fallout. The Economist

United Nations Economic Commission for Europe (2017) Implementation of the Convention on Environmental Impact Assessment in a Transboundary Context (2013-2015) Fifth review. United Nations Publication

U.S. FDA, Center for Veterinary Medicine (2017) Guidance for Industry #187, Regulation of Intentionally Altered Genomic DNA in Animals, Draft Guidance.

U.S. Nuclear Regulatory Commission (2004) Nuclear Power Plant Licensing Process.

U.S. Senate Committee on Environment and Public Works Majority Staff (2006) Yucca Mountain: The most studied real estate on the planet.

Whipple C (2006) Performance Assessment: What Is It and Why Is It Done? In: Macfarlane A, Ewing RC (eds) Uncertainty Underground: Yucca Mountain and the Nation's High-level Nuclear Waste. MIT Press, Cambridge, MA,

Wong E (2016) Coal Burning Causes the Most Air Pollution Deaths in China, Study Finds. The New York Times

Zhang S (2017) The White House Revives a Controversial Plan for Nuclear Waste. The Atlantic